\documentclass[nofootinbib, pre, twocolumn]{revtex4}

\usepackage{amsmath}
\usepackage{amssymb}
\usepackage{graphicx}
\usepackage{bm}
\usepackage{braket}
\usepackage{color, soul}
\usepackage[dvipsnames]{xcolor}
\usepackage{bbm}
\DeclareMathOperator{\id}{\mathbbm{1}}
\graphicspath{ {./} }
\usepackage{float}
\usepackage{soul}
\usepackage{capt-of}

\definecolor{JM}{RGB}{4,116,149}

\newcommand{\beq}{\begin{equation}}
\newcommand{\eeq}{\end{equation}}

\begin{document}

\title{Point Processes with Gaussian Boson Sampling}
\author{Soran Jahangiri}
\affiliation{Xanadu, 777 Bay Street, Toronto, Canada}
\author{Juan Miguel Arrazola}
\affiliation{Xanadu, 777 Bay Street, Toronto, Canada}
\author{Nicol\'as Quesada}
\affiliation{Xanadu, 777 Bay Street, Toronto, Canada}
\author{Nathan Killoran}
\affiliation{Xanadu, 777 Bay Street, Toronto, Canada}

\begin{abstract}
\vspace{0.1cm}
Random point patterns are ubiquitous in nature, and statistical models such as point processes, i.e., algorithms that generate stochastic collections of points, are commonly used to simulate and interpret them. We propose an application of quantum computing to statistical modeling by establishing a connection between point processes and Gaussian Boson Sampling, an algorithm for special-purpose photonic quantum computers. We show that Gaussian Boson Sampling can be used to implement a class of point processes based on hard-to-compute matrix functions which, in general, are intractable to simulate classically. We also discuss situations where polynomial-time classical methods exist. This leads to a family of efficient quantum-inspired point processes, including a new fast classical algorithm for permanental point processes. We investigate the statistical properties of point processes based on Gaussian Boson Sampling and reveal their defining property: like bosons that bunch together, they generate collections of points that form clusters. Finally, we discuss several additional properties of these point processes which we illustrate with example applications.\\
\end{abstract}

\maketitle

\section{Introduction}
Despite their stochastic nature, quantum algorithms have often been studied in contexts where their intrinsic randomness is an obstacle rather than a benefit. For instance, consider Shor's factoring algorithm: it constructs states such that, with high probability, their measurement outcomes can be post-processed to reveal the prime factors of an input composite number \cite{shor1994algorithms}. The quantum computer is acting as a sampler whose probability distribution is highly concentrated on outcomes that reveal the solution to the factoring problem. However, it would be preferable to obtain the desired answers with certainty rather than with high probability. Other quantum algorithms can also be viewed in this light: Grover's search algorithm samples outputs that are likely to contain a marked item in an unstructured database \cite{grover1997quantum}; the quantum algorithm for linear systems of equations randomly outputs large elements of the solution vector \cite{harrow2009quantum}; and the quantum approximate optimization algorithm reveals bit strings that have a high chance of being good approximations to the solution of combinatorial optimization problems \cite{farhi2014quantum}. Nevertheless, randomness in quantum algorithms can be harnessed and turned into a feature when applied to the right problems. \\

Stochastic processes occur in abundance in nature as well as in human affairs, and understanding them requires building models that can reproduce their unique random properties. This is, in essence, the goal of statistical modeling: to build accurate mathematical representations of random processes \cite{freedman2009statistical}. Point processes are statistical models that generate random collections of data points according to a given probability distribution \cite{cox1980point, daley2003introduction, daley2007introduction, baddeley2007spatial}. Similarly, point process analysis is a quantitative statistical method for analyzing such point patterns to provide information that can be used for prediction and planning purposes \cite{baddeley2006case}. This finds applications in a variety of fields such as finance \cite{bjork1997bond, bauwens2009modelling, embrechts2011multivariate, bacry2015hawkes}, seismology \cite{ogata1988statistical, ogata1998space, kagan1977earthquake}, biology \cite{Li2016Point, Fromion2013Stochastic, Dodgson2013Spatial}, medicine \cite{Jing2010Detection, Grell2015Three, Fok2012Functional}, ecology \cite{Illian2017Improving, Igea2019Multiple, Warton2010Poisson, Mi2014Point, Illian2013Fitting}, physics \cite{Picinbono2010Some, Klatt2014Characterization, Wohrer2019Ising, Schmidt2017Inertial}, and chemistry \citep{Halle2013Analysis, Persson2013Transient, Tzoumanekas2006Topological, Lazenby2018Quantitative, Talattof2018Pulse}.\\ 

Several point processes are based on probability distributions that select points according to matrix functions. An example is the determinantal point process (DPP) \cite{macchi1975coincidence,borodin2009determinantal,lavancier2015determinantal}, which, as the name suggests, is based on the determinant as the underlying matrix function. DPPs have been studied in depth by the mathematics community \cite{borodin2000distributions,shirai2000fermion, okounkov2001infinite,okounkov2003correlation, borodin2003janossy, johansson2004determinantal, johansson2005random,borodin2005eynard, borodin2010adding, hough2006determinantal} and have found a large number of applications, notably in machine learning \cite{kulesza2010structured, kulesza2011k, lin2012learning,kulesza2012determinantal,gillenwater2014expectation} and optimization \cite{hartline2008optimal, dughmi2009revenue, lee2010diverse, kim2011distributed, affandi2013nystrom}. The wide adoption of DPPs is partially due to the fact that determinants can be efficiently calculated, leading to fast implementations of DPPs. The same is not true for other matrix point processes: even if they are of great potential interest, they find limited usage due to the hardness of their numerical deployment. \\ 

In this work, we introduce a new class of point processes -- the hafnian and Torontonian point processes (HPPs and TPPs) -- and show that they can be natively implemented using a special-purpose photonic quantum algorithm known as Gaussian Boson Sampling (GBS). The hafnian is a generalization of the permanent, and indeed HPPs contain permanental point processes \cite{mccullagh2006permanental, kogan2010permanental} as a special case. While DPPs generate points that are scattered -- like fermions obeying the Pauli exclusion principle -- HPPs and TPPs sample points that are clustered -- like bosons that bunch together. In the most general case, implementing these point processes using classical methods cannot be done in polynomial time. However, for specific instances, efficient classical simulation algorithms exist, which give rise to a new class of quantum-inspired point processes. We demonstrate the clustering property of these point processes through qualitative and quantitative analyses, and explore their usefulness by studying potential applications.  

\clearpage

\section{Background}

For completeness, we provide a brief background of both point processes and GBS.

\begin{figure}
\includegraphics[scale=0.65]{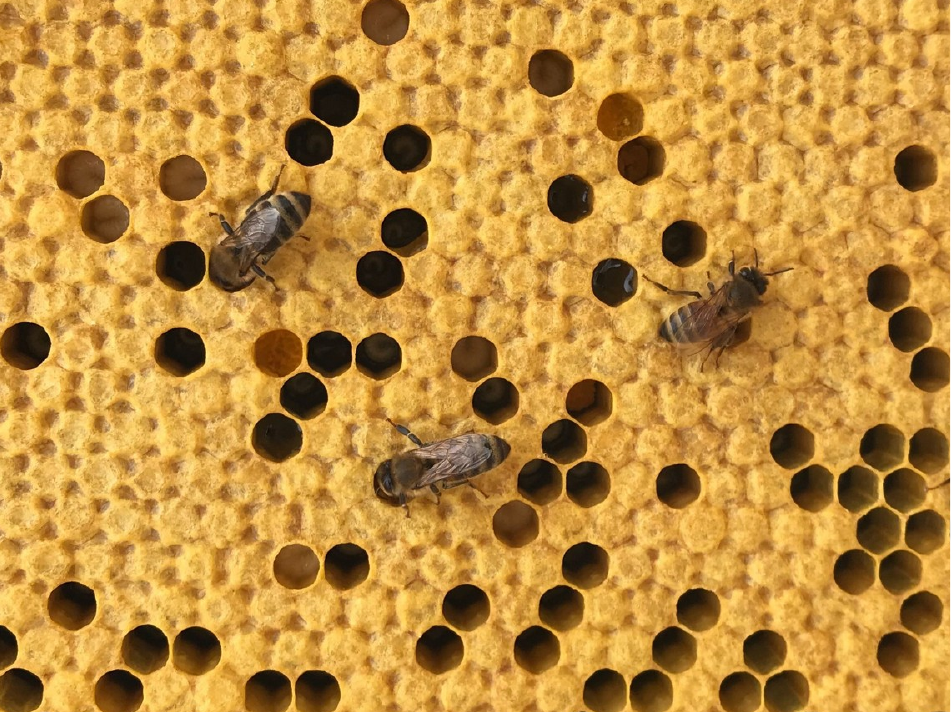}
\centering
\caption{An illustration of a random point pattern in a honeycomb. The sealed yellow cells contain developing bees and the dark holes are empty cells that have been randomly vacated. The appearance of empty cells can be modeled as a point process generating clusters of points in the hexagonal lattice.}\label{fig:bee}
\end{figure}

\vspace{-0.3cm}

\subsection{Point Processes}\label{Sec:PP}
Random point patterns occur ubiquitously in nature and human affairs. This is illustrated for example in Fig.~\ref{fig:bee}, where we show a random pattern of empty cells in a honeycomb. Point processes provide a method to model and analyze these random patterns as well as the mechanisms that underlie them. Informally, a point process is a mechanism that randomly generates points among a set of possible outcomes. More precisely, a point process $\mathcal{P}$ is a probability distribution over subsets of a given state space $\mathcal{M}$. We focus on discrete point processes, in which case the state space can be enumerated by a finite set $\mathcal{M} = \{1, . . . , m\}$ containing $m$ members, and the distribution $\mathcal{P}$ is defined over the power set $2^{\mathcal{M}}$, the set of all $2^m$ possible subsets of $\mathcal{M}$. Each subset of $\mathcal{M}$ is denoted by $S\subseteq\mathcal{M}$. A point process is thus uniquely specified by the choice of distribution $\mathcal{P}$ and state space $\mathcal{M}$.\\

In a uniformly random point process, $\mathcal{P}$ is a uniform distribution over all subsets of $\mathcal{M}$. When the number of points in such a random point process is restricted to follow a Poisson distribution with a constant intensity rate, the resulting point process is a homogeneous Poisson point process (PPP). In a PPP, the probability $\mathcal{P}$ is uniform over all outputs with an equal number of points. A PPP can be considered as a point process with no interaction between the points and therefore is likely to contain both local clusters and voids. The term interaction here refers to the effect that a point at a specific location might have on the appearance of points in other locations. A natural way to generate other types of point processes is to introduce some level of interaction between the points, which may favor clustering or scattering. An important class of such point processes are formed by relating the probability of observing a particular output pattern to matrix functions such as determinants and permanents.\\

We define a \emph{matrix point process} as a point process where the probability $\mathcal{P}(S)$ of observing an outcome $S$ takes the form
\beq
\mathcal{P}(S)=\frac{\Phi(K_{S})}{\mathcal{N}},
\eeq
where $K$ is an $m\times m$ symmetric \emph{kernel matrix}, $K_S:=[K_{i,j}]_{i,j\in S}$ is a submatrix of $K$ obtained by keeping rows and columns corresponding to the outcome $S$, $\Phi$ is a matrix function, with both $\Phi$ and $K$ chosen to ensure that $\Phi(K_S)\geq 0$ for all $S\in 2^{\mathcal{M}}$, and $\mathcal{N}$ is a normalization constant. Specific classes of matrix point processes are determined by the choice of matrix function and, among each class, the properties of the resulting point process are uniquely determined by the choice of kernel matrix. For instance, a determinantal point process (DPP) is defined by the distribution~\cite{kulesza2012determinantal}  
\begin{equation}
\mathcal{P}(S) = \frac{\mathrm{det}(K_S)}{\mathrm{det}(K + \id)},
\end{equation}
where $\id$ is the identity matrix. Instead of the full probability distribution, it is often more convenient to work with the $n$-point correlation function of matrix point processes. For a given subset $\mathbf{r}=(r_1, \ldots, r_n)\subseteq \mathcal{M}$ with $n\leq m$, it is defined as~\cite{Cunden2019Free}
\begin{equation} \label{Eq: rho_n}
\rho_n(r_1, \ldots, r_n) =  \Phi(K_{r_i,r_j})_{i,j=1,\ldots,n}.
\end{equation} 
The correlation function is the unnormalized probability of observing the output $(r_1, \ldots, r_n)$ appearing among the elements of a sample $S$ drawn from $\mathcal{P}$. Therefore, it quantifies the likelihood that these points appear together when generating samples from the matrix point process. For example, the correlation function of a DPP for two points in the state space, referred to as the 2\emph{-point correlation function}, is given by
\begin{equation} \label{Eq: dpp_rho2}
\rho_2(r_1, r_2) = K_{r_1,r_1}K_{r_2,r_2} - (K_{r_1,r_2})^2,
\end{equation} 
where we have used the fact that the kernel matrix is symmetric. If $K$ measures similarity between points, namely if $K_{r_i,r_j}$ takes large values when $r_i$ and $r_j$ are similar to each other, Eq.~\eqref{Eq: dpp_rho2} shows that similar points are unlikely to occur together, i.e., DPPs lead to diversification.   

\vspace{-0.3cm}

\subsection{Gaussian Boson Sampling}\label{Sec: GBS}
In a quantum-optical setting, the state of a system of $m$ optical modes can be uniquely specified by its so-called Wigner function $W(\bm{q},\bm{p})$ \cite{weedbrook2012gaussian,serafini2017quantum}, where $\bm{q} \in \mathbb{R}^m$ are the so-called canonical positions and $\bm{p} \in \mathbb{R}^m$ are the canonical momenta of the state. Gaussian states are the set of quantum states with Gaussian Wigner functions. Just like multidimensional Gaussian distributions, Gaussian states are specified by a covariance matrix $V$ and a vector of means $\bm{\bar{q}},\bm{\bar{p}}$. Besides being a positive definite covariance matrix, a valid quantum covariance matrix must satisfy the uncertainty principle 
\begin{align}\label{eq:uncertaintyxp}
V + i \tfrac{\hbar}{2} \Omega \geq 0,
\end{align}
where $\Omega = \left( \begin{smallmatrix}
0 & \mathbb{I}_m \\
-\mathbb{I}_m &0
\end{smallmatrix} \right)$ is the symplectic matrix and $\hbar$ is a positive constant. It will be convenient to write the covariance matrix in terms of the complex amplitudes $\bm{\alpha} = \tfrac{1}{\sqrt{2 \hbar }} (\bm{q}+i \bm{p}) \in \mathbb{C}^m$. The variables $\bm{\alpha}$ are said to be complex normal distributed with mean $\bm{\bar{\alpha}} = \tfrac{1}{\sqrt{2 \hbar}} (\bm{\bar{q}}+i \bm{\bar{p}}) \in \mathbb{C}^m$ and covariance matrix ${\Sigma}$  \cite{picinbono1996second}, which furthermore also needs to satisfy the uncertainty relation \cite{simon1994quantum}
\begin{align}\label{eq:uncertainty}
{\Sigma} + {Z}/2 \geq 0,
\end{align}
where ${Z} = \left( \begin{smallmatrix}
\mathbb{I}_m & 0\\
0& -\mathbb{I}_m 
\end{smallmatrix} \right)$.
The covariance matrix $\Sigma$ is customarily parameterized as \cite{picinbono1996second}
\begin{align}
\Sigma = \left(\begin{array}{c|c}
\Gamma & C \\
\hline
C^* & \Gamma^*
\end{array} \right),
\end{align}
where $C$ is symmetric and $\Gamma$ is hermitian and positive definite.\\

Gaussian Boson Sampling (GBS) is a model of photonic quantum computing where a Gaussian state is measured using photon-number-resolving detectors. A general Gaussian state can be prepared by using single-mode squeezing and displacement operations together with linear-optical interferometry. It was shown in Ref.~\cite{hamilton2017gaussian} that when the modes of a Gaussian state with zero mean ($\bm{\alpha} = 0$) are measured, the probability of obtaining a pattern of photons $S = (s_1,\ldots,s_m)$, where $s_i$ is the number of photons in mode $i$, is given by
\begin{align}\label{eq:GBS}
\mathcal{P}(S) = \frac{1}{\sqrt{\text{det}(\sigma_Q)} } \frac{\text{Haf}(K_S)}{s_1!\ldots s_m!}, 
\end{align}
where 
\begin{align}
\sigma_Q&:=\Sigma +\mathbb{I}_{2m}/2,\\
K &:= X \left(\mathbb{I}_{2m} - \sigma_Q^{-1}\right)\label{Eq:A_matrix},
\end{align}
and $K_S$ is the matrix obtained by repeating columns and rows $i$ and $i+M$ of the kernel matrix $K$ a number of times equal to $s_i$. If $s_i=0$ then the rows and columns $i$ and $i+M$ are deleted from $K$ in order to form $K_S$. The matrix function ${\rm Haf}(\cdot)$ is the 
\emph{hafnian} \cite{caianiello1953quantum} which, for a $2m\times 2m$ matrix $K$, is defined as
\begin{equation}
{\rm Haf}(K) = \sum_{\mu\in {\rm PMP}} \prod_{(i,j)\in \mu} K_{i,j},
\end{equation}
where $\rm PMP$ is the set of perfect matching permutations, namely 
the possible ways of partitioning the set $\{1,\dots,2m\}$ into subsets of size 2. The hafnian is \#P-Hard to approximate for worst-case instances  \cite{barvinok2016combinatorics} and the runtime of the best known algorithms for computing hafnians scales exponentially with the dimension of the input matrix \cite{bjorklund2018faster}. The difficulty of computing the hafnian has been used to show that sampling from general GBS distributions cannot be done in classical polynomial time unless the polynomial hierarchy collapses \cite{aaronson2011computational,hamilton2017gaussian}.\\

In GBS, it is possible that more than one photon can be observed in a given output mode, i.e., it is possible that $s_i>1$. In certain cases, only the location of the detected photons is relevant --- not their quantity at a given mode --- so it becomes convenient to set $s_i=1$ for any $s_i>1$. Physically, this is precisely the effect of threshold detectors: they `click' whenever one or more photons are observed. It was shown in Ref.~\cite{quesada2018gaussian} that the resulting GBS distribution when employing threshold detectors is given by
\beq\label{Eq: Tor_dbn}
\mathcal{P}(S)=\frac{1}{\sqrt{\det(\sigma_Q)}}\text{Tor}\left( X K_S\right),
\eeq
where $X=\begin{pmatrix}
0 & \id_{|S|}\\
\id_{|S|} & 0
\end{pmatrix}$ and $\text{Tor}(\cdot)$ is the \emph{Torontonian}, which for a $2m\times 2m$ matrix $K$ is defined as 
\beq
\text{Tor}(K)=\sum_{Z\in 2^{\mathcal{M}}}(-1)^{|Z|}\frac{1}{\sqrt{\det(\id-K_Z)}},
\eeq
where $\mathcal{M}=\{1,2,\ldots,m\}$ and $2^{\mathcal{M}}$ denotes its powerset.

\section{Point Processes with Gaussian Boson Sampling}

\begin{figure*}[t!]
\centering
\includegraphics[scale=0.8]{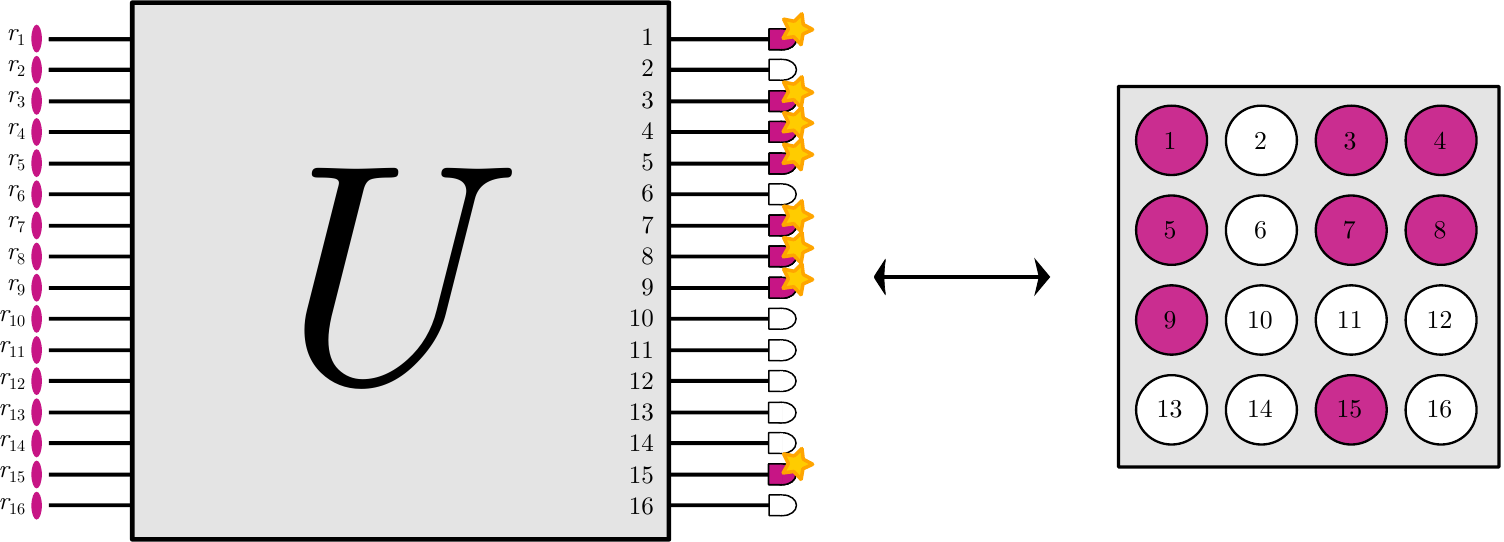}
\caption{Schematic illustration of a point process implemented with GBS. A symmetric kernel matrix $K$ can be encoded by appropriately selecting the squeezing levels $r_1, r_2, \ldots, r_m$ and the interferometer unitary $U$. Each point in the state space is associated with an output mode such that the modes where photons are observed determine the specific point pattern that has been sampled.}\label{fig:GBS_scheme}
\end{figure*}

Once the mathematical concepts of point processes and GBS have been placed alongside each other, their connection is evident: a GBS device is a physical realization of a matrix point process. A schematic illustration of this connection is shown in Fig.~\ref{fig:GBS_scheme}. In this section, we make that connection explicit and analyze the properties of the resulting point process.\\ 

For any positive integer $m$, consider a state space consisting of vectors $(s_1,s_2,\ldots,s_m)$ such that each entry $s_i$ is a non-negative integer and the sum of all entries is an even number, i.e., $\sum_{i=1}^n s_i\mod 2 = 0$. We define a \emph{hafnian point process} (HPP) by the probability distribution
\begin{equation} \label{Eq: pgbs}
\mathcal{P}(S) = \frac{1}{\sqrt{\det(\sigma_Q)}}\frac{\text{Haf}(K_S)}{s_1!\ldots s_m!},
\end{equation}
where $K$ is a $2m\times 2m$ symmetric kernel matrix. An HPP is therefore simply a matrix point process with the hafnian as the matrix function.\\

The hafnian is a generalization of the permanent, in the sense that the permanent of an arbitrary matrix $K$ can be expressed in terms of the hafnian of a related matrix using the identity
\begin{equation} \label{Eq: HafPer}
\mathrm{per}(K) = \mathrm{Haf} 
\begin{pmatrix}
0 & K \\
K^T & 0 \\
\end{pmatrix}.
\end{equation}
Consequently,  HPPs generalize permanental point processes: they contain them as a special case. Similarly, for a state space $\mathcal{M}=\{1,2,\ldots,m\}$, we define a \emph{Torontonian point process} (TPP) by the distribution
\beq
\mathcal{P}(S) = \frac{\text{Tor}(K_S)}{\sqrt{\det(\sigma_Q)}}.
\eeq
We refer to both HPPs and TPPs as GBS point processes. We now study sufficient conditions to embed specific types of kernel matrices into a GBS device.

\subsection{Kernel matrices}

Williamson's theorem \cite{williamson1936algebraic} combined with the Bloch-Messiah reduction \cite{bloch1975canonical} provide a recipe to prepare an arbitrary Gaussian quantum state by using combinations of single-mode squeezing, single-mode displacements, and linear optics interferometers \cite{serafini2017quantum}. In general, to encode a given matrix into a GBS device, the conditions are that the covariance matrix ${\Sigma}$ satisfies the uncertainty relation of Eq.~\eqref{eq:uncertainty} and that the kernel matrix $K$ is defined in terms of $\Sigma$ according to Eq.~\eqref{Eq:A_matrix}.  \\

Following Ref.~\cite{Rahimi2015What}, we consider circuits that take as inputs single-mode Gaussian states characterized by a diagonal covariance matrix $V_i = \text{diag}\left( V_q^{(i)},V_p^{(i)} \right)$ in the $q_i,p_i$ variables. In order to satisfy the uncertainty relation in Eq.~\eqref{eq:uncertaintyxp} the variances must satisfy 
\begin{align}\label{eq:uncertaintyv}
V_q^{(i)} V_p^{(i)} \geq (\hbar/2)^2.
\end{align}
A state that has both $V_q^{(i)} \geq \hbar/2$ and $V_p^{(i)} \geq \hbar/2$ will be termed ``classical", since both of its variances have fluctuations above the noise of the vacuum state, $\hbar/2$.\\

Having prepared the inputs, single-mode Gaussian states are sent through a linear-optical interferometer, which physically corresponds to an array of half-silvered mirrors and waveplates and enables generation of entanglement between the different modes. Mathematically, an interferometer can be uniquely described by a unitary matrix $U$ with dimension equal to the number of modes. With this description we can explicitly construct the covariance matrix of the output or, more interestingly, we can directly construct the kernel matrix $K$ appearing in Eq.~\eqref{Eq:A_matrix}, which is given by:
\begin{align}\label{defA}
{K} =  \left[	
\begin{array}{c|c}
{B} & {C} \\
\hline
{C}^* & {B}^* \\
\end{array}
\right] = {K}^T,
\end{align}
where
\begin{align}
{C} &= {U}\, \text{diag}(\mu_1,\mu_2,\ldots,\mu_m)	\,{U}^\dagger = {C}^\dagger, \label{Eq: C matrix}\\
{B} &= {U} \,\text{diag}(\lambda_1,\lambda_2,\ldots,\lambda_m) \,	{U}^T = {B}^T, \\
 \mu_i = 1-&\left(\frac{1}{1+2V_q^{(i)}/\hbar}+\frac{1}{1+2V_p^{(i)}/\hbar} \right),\\\lambda_i = &\frac{1}{1+2V_p^{(i)}/\hbar}-\frac{1}{1+2V_q^{(i)}/\hbar}.
\end{align}
Now we consider in detail certain choices of input states and the resulting kernel matrices.

\subsubsection{Squeezed states}\label{Sec:sq}
For squeezed states with squeezing level $r$, one of the quadratures, say $V_p^{(i)} = \tfrac{\hbar}{2}e^{-2r_i}$, is squeezed below the vacuum level while the other quadrature is antisqueezed by exactly the opposite amount $V_q^{(i)} = \tfrac{\hbar}{2}e^{2r_i}$. Under these circumstances it holds that $\mu_i=0$, $\lambda_i=\tanh(r_i)$, and
\begin{align}
{K}_{\text{sq}} =  \left[	
\begin{array}{c|c}
{B} & {0} \\
\hline
{0} & {B}^* \\
\end{array}
\right] .
\end{align}	
The matrix ${B}$ can be an arbitrary symmetric matrix except for the restriction that its singular values must satisfy $\lambda_i = \tanh(r_i) \in [0,1)$. To see this, consider the Takagi-Autonne decomposition \cite{cariolaro2016bloch,caves2017,horn1990matrix} of any complex symmetric matrix, given by
\begin{align}
{B} = {U}\, \text{diag}(\lambda_1, \lambda_2,\ldots, \lambda_m)\, {U}^T	.
\end{align}
The values $\lambda_i \geq 0$ are the singular values of ${V}$ and the Takagi-Autonne decomposition is therefore just a fine-tuned version of the singular value decomposition for symmetric matrices. Assuming $\lambda_i <1$ then we can always write $\tanh(r_i) = \lambda_i$.\\

In the case of squeezed input states, the GBS probability distribution satisfies
\begin{equation} \label{Eq: pgbs_squared}
\mathcal{P}(S) = \frac{1}{\sqrt{\det(\sigma_Q)}}\frac{|\text{Haf}(B_S)|^2}{s_1!\ldots s_m!}.
\end{equation}
This formula therefore states that hafnian point processes can be designed for any complex symmetric kernel matrix $B$.\\

The average number of points generated by a GBS point process can be controlled by suitably rescaling the kernel matrix: $K\rightarrow c K$, where $c>0$ is a constant. Letting $N$ denote the number of points generated, the average number of points $\mathbb{E}(N)$ satisfies
\beq
\mathbb{E}(N)=\sum_{i=1}^m\frac{(c\,\lambda_i)^2}{1-(c\, \lambda_i)^2},
\eeq
which can be solved for $c$ to set any desired average number of points.

\subsubsection{Thermal states}\label{Sec:th}
Single-mode thermal states are characterized by a covariance matrix satisfying $V_p^{(i)}=V_q^{(i)} = \hbar \left(\bar{n}_i+\tfrac{1}{2} \right)$, where $\bar{n}_i$ is the mean photon number of mode $i$. For these states we also have $\mu_i = \frac{\bar{n}_i}{1+\bar{n}_i}$, $\bar{n}_i\geq 0$, $\lambda_i=0$, and 
\begin{align}
{K}_{\text{th}} &=  \left[	
\begin{array}{c|c}
{0} & {C} \\
\hline
{C}^T & {0} \\
\end{array}
\right],\\
C = U\,& \text{diag}(\mu_1, \mu_2,\ldots, \mu_m)\, U.
\end{align}	
Except for proportionality constants, the matrix $C$ can be made proportional to an arbitrary positive semidefinite complex matrix since $\mu_i\geq 0$ in Eq.~\eqref{Eq: C matrix}. Finally, using Eq.~\eqref{Eq: HafPer}, we conclude that a GBS device with thermal states as input can be used to sample from a permanental point process characterized by the distribution
\beq\label{Eq:P(S)_thermal}
\mathcal{P}(S) = \frac{1}{\sqrt{\det(\sigma_Q)}}\frac{\text{per}(C_S)}{s_1!\ldots s_m!},
\eeq
where $C$ is an arbitrary positive semidefinite kernel matrix. The average number of points in this case satisfies
\beq
\mathbb{E}(N)=\sum_{i=1}^m\frac{c \mu_i}{1-c\mu_i},
\eeq
where $c$ is the rescaling constant of the kernel matrix.
  
\subsubsection{Squashed states}\label{Sec:squ}
Squashed states are single-mode states with the property that the variance in both position and momentum are above vacuum fluctuations, i.e., $V_q^{(i)}, V_p^{(i)} \geq \hbar/2$ and $V_p^{(i)} \neq V^{(i)}_q$. They differ from squeezed states, where one quadrature has below-vacuum fluctuations, and from thermal states, where $V_p^{(i)} = V^{(i)}_q$. We consider the specific situation where $V_q^{(i)}=\hbar/2$ for all $i$, in which case
\begin{align}
\mu_i  = \lambda_i = \frac{1}{2}-\frac{1}{1+2 V_p^{(i)}/\hbar}.	
\end{align}
We parameterize $V^{(i)}_p = \tfrac{\hbar}{2}\exp(2 r_i)$ with $r_i \geq 0$ and restrict the interferometer such that it is characterized by a real orthogonal matrix $O$. We then have:
\begin{align}
{C}= {B} &= \tfrac{1}{2}{O} \  \text{diag}(\lambda_1, \lambda_2,\ldots,\lambda_m)  \ 	{O}^T,\\
&{K}_{\text{sqsh}} =  \left[	
\begin{array}{c|c}
{C} & {C} \\
\hline
{C} & {C} \\
\end{array}
\right].
\end{align}
Except for proportionality constants, the matrix $C$ can be chosen to be an arbitrary positive semidefinite real matrix. This gives rise to an HPP with probability distribution
\beq
\mathcal{P}(S) = \frac{1}{\sqrt{\det(\sigma_Q)}}\frac{\text{Haf}({K}_{\text{sqsh},S})}{s_1!\ldots s_m!}.
\eeq
The average number of points satisfies
\beq
\mathbb{E}(N)=\sum_{i=1}^m\frac{(2c\lambda_i)}{1-(2c\lambda_i)},
\eeq
where as before, $c$ is the rescaling constant of the kernel matrix.

\subsection{Quantum-inspired point processes}\label{Sec:GBSi}

In general, the probability distribution of Eq.~\eqref{Eq: pgbs} cannot be sampled from in classical polynomial time, in which case photonic quantum devices are needed to implement GBS point processes. Nevertheless, as we now show, for kernel matrices satisfying specific properties, the resulting point processes can be implemented in polynomial time using classical computers. The resulting algorithms are efficiently-implementable quantum-inspired classical point processes. The main idea is that classical Gaussian states, i.e., states whose variances are above vacuum level for both quadratures, can be represented in terms of probability distributions over states whose interaction through linear-optical interferometers can be straightforwardly computed. \\ 

Similarly to the Wigner function, any single-mode quantum state $\tau$ can be uniquely represented in terms of the so-called $P$ representation
\begin{align}\label{Eq:P(alpha)}
\tau = \int_{\mathbb{C}} d^2 \alpha \ P(\alpha) |\alpha \rangle \langle \alpha|,
\end{align}
where $|\alpha \rangle \langle \alpha|$ represents a coherent state with parameter $\alpha$. Coherent states are Gaussian states with variances $V_q^{(i)} = V_p^{(i)} =\hbar/2$ and complex amplitude $\alpha$. The function $P(\alpha)$ is a quasi-probability distribution in the sense that it can take negative values, but there exist states for which it is positive over its entire domain. In such cases, the right-hand side of Eq.~\eqref{Eq:P(alpha)} can be interpreted as a probability distribution over coherent states. Following a result of Ref.~\citep{Rahimi2015What}, if a single-mode Gaussian state $\tau$ is classical, i.e., with variances $V_p^{(i)},V_q^{(i)} \geq \hbar /2$, then it has a positive $P$ function. This includes thermal states and squashed states, whose $P$ functions are respectively given by
\begin{align}
&P_{\text{th}}(\alpha) = \frac{1}{\pi \bar{n}}\exp\left(-\frac{\alpha_R^2+\alpha_I^2}{\bar{n}}\right),\label{Eq:P_th}\\
P_{\text{sqsh}}(\alpha)&=\frac{\delta(\alpha_R)}{\sqrt{\pi (e^{2r}-1)/2}} \exp\left(-\frac{\alpha_I^2}{(e^{2r}-1)/2}\right) , \quad r \geq 0\label{Eq:P_sqsh},
\end{align}
where $\alpha_R,\alpha_I$ are respectively the real and imaginary parts of $\alpha$ and $\delta(\alpha)$ is the Dirac delta function. States with a positive $P$ representation can be prepared by sampling the random variable $\alpha$ with probability density function $P(\alpha)$ and then preparing the resulting state $\ket{\alpha}\bra{\alpha}$. If the inputs of an $m$-mode interferometer characterized by a unitary $U$ are independent coherent states with parameters $\alpha_1,\alpha_2,\ldots,\alpha_m$, the output states are also independent coherent states with parameters $\beta_1,\beta_2,\ldots,\beta_m$, where
\beq
\beta_i =\sum_{j=1}^m U_{ji} \alpha_j.
\eeq
Finally, the photon number distribution of a coherent state with parameter $\beta$ is a Poisson distribution with mean $|\beta|^2$, so a sample photon pattern can be obtained by sampling each mode independently according to its Poisson distribution.\\

In summary, the classical sampling algorithm for Gaussian states with positive $P$ representation works as follows:

\begin{figure*}[t!]
\centering
\includegraphics[scale=0.35]{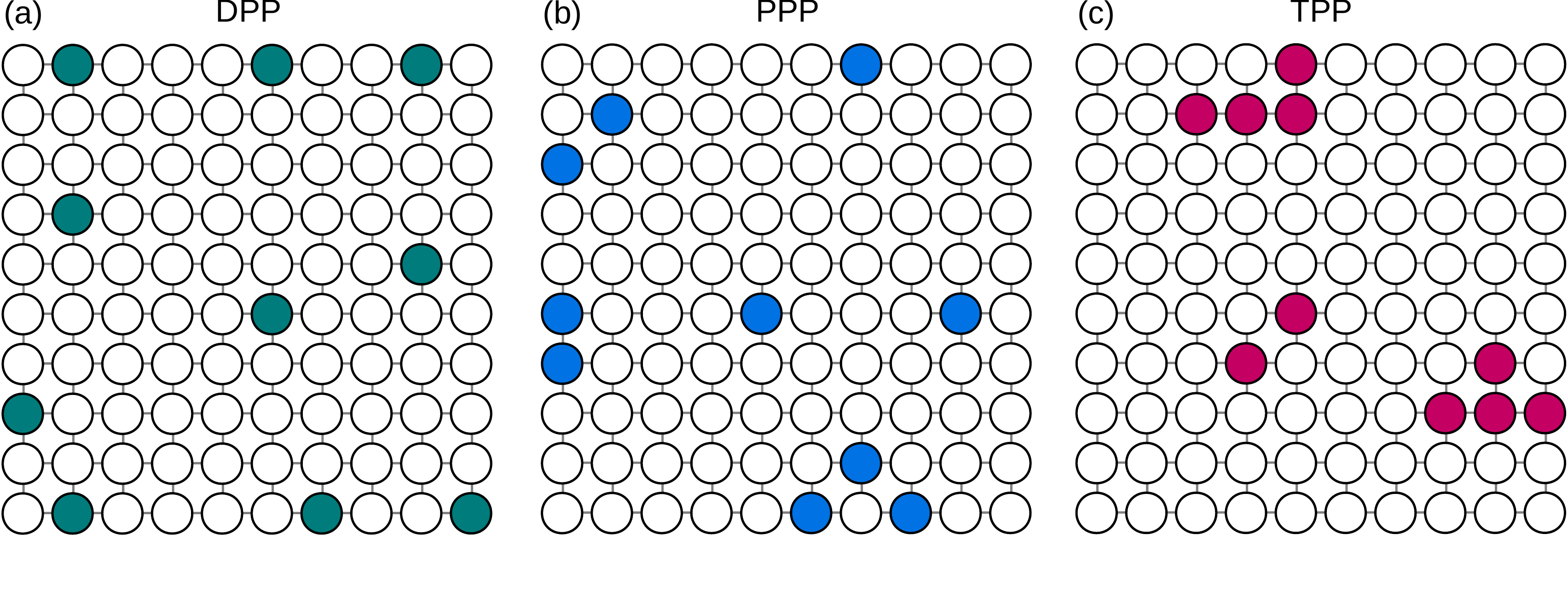}
\caption{Samples generated with (a) determinantal, (b) Poisson, and (c) Torontonian spatial point processes in a two-dimensional space containing 100 points. DPP samples have points that are scattered and spread out in space. PPPs treat all patterns uniformly, so both clustering and repulsion are typically present. For TPPs, sample points are likely to occur in clusters.}\label{fig:pdh}
\end{figure*}

\begin{figure}[b]
\centering
\includegraphics[scale=0.39]{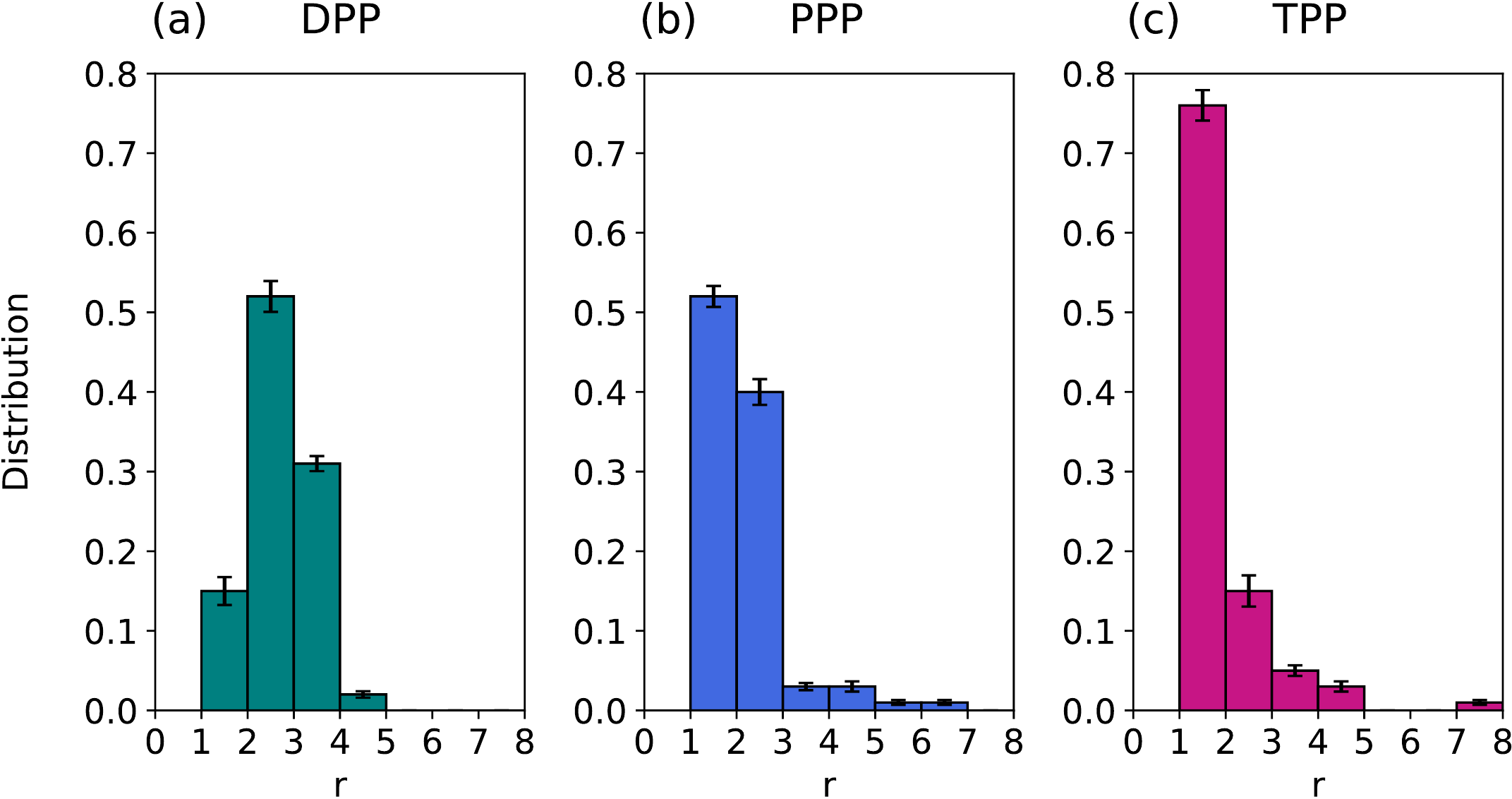}
\caption{Normalized distributions of the nearest-neighbour distance for (a) determinantal, (b) Poisson, and (c) Torontonian spatial point processes in a two-dimensional space containing 100 points. We set a value of $\sigma=1$ for the kernel matrix parameter in Eq.~\eqref{Eq:K_ij}. The statistics were taken over ten independent samples and the error bars correspond to one standard deviation. For DPPs, the most common nearest-neighbour distance is noticeably larger than those for PPPs and TPPs, showcasing the scattering property of DPPs. Conversely, the significant majority of TPP points have their nearest-neighbour at the closest possible distances, a signal of the clustering property of this point process.}\label{fig:nnd}
\end{figure}

\begin{enumerate}
\item For each mode $i=1,2,\ldots, m$ with input state $\tau_i = \int_{\mathbb{C}} d^2 \alpha \ P_i(\alpha_i) |\alpha_i \rangle \langle \alpha_i|$, sample $\alpha_i$ according to the distribution $P_i(\alpha_i)$. 
\item For each mode, compute the output parameter $\beta_i =\sum_{j=1}^m U_{ji} \alpha_j$ and sample the photon number $s_i$ from a Poisson distribution with mean $|\beta_i|^2$.
\item Return the sample $S=(s_i, s_2,\ldots, s_m)$.\\
\end{enumerate}
For thermal and squashed states, the distributions $P(\alpha)$ of Eqs.~\eqref{Eq:P_th} and \eqref{Eq:P_sqsh} are Gaussian in the parameters $\alpha_R,\alpha_I$, so this sampling can be done in $O(1)$ time for each mode. Similarly, sampling from Poisson distributions can be done in $O(1)$ for each mode. The overhead of the algorithm arises from the complexity of computing the parameters $\beta$, which in total takes only $O(m^2)$ time. Importantly, as shown in Eq.~\eqref{Eq:P(S)_thermal}, GBS with thermal states gives rise to a permanental point process for positive semidefinite and real kernel matrices. Our method is thus an efficient algorithm for implementing these permanental point processes, running in $O(m^2)$ time. By contrast, known algorithms for implementing determinantal point processes \cite{kulesza2012determinantal} rely on diagonalization of the $m\times m$ kernel matrix, which takes $O(m^3)$ time.

\section{Properties of GBS point processes}\label{Sec:prop}
\begin{figure*}[t!]
\includegraphics[scale=0.39]{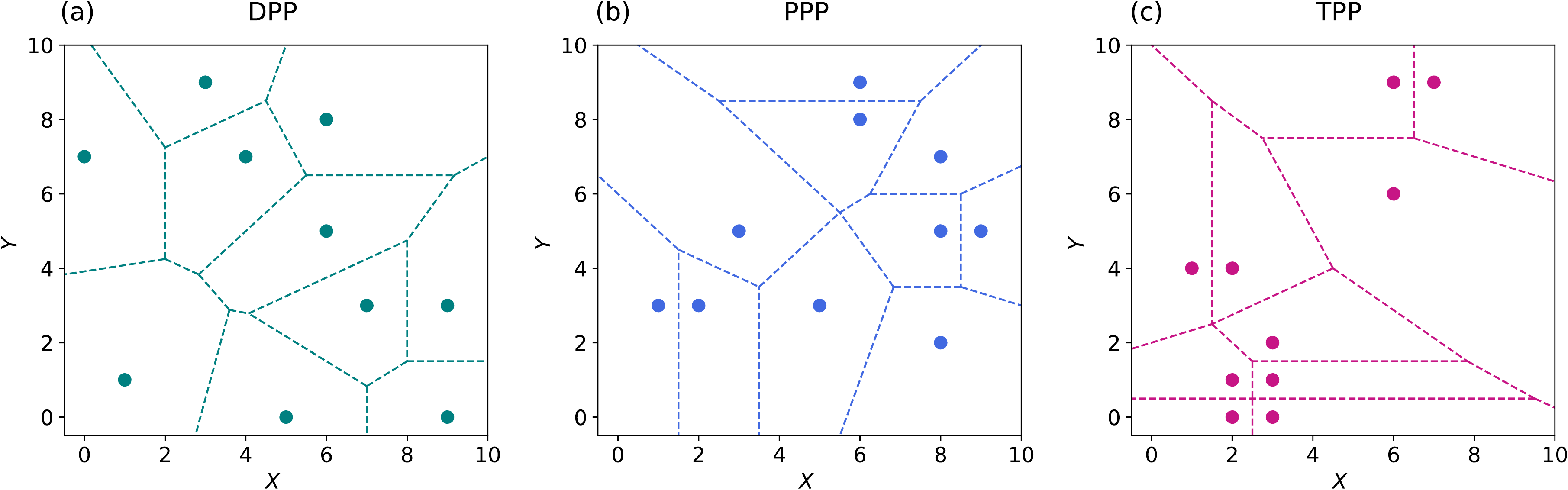}
\caption{Examples of Voronoi cell diagrams for points obtained from (a) determinantal, (b) Poisson, and (c) Torontonian point processes in a two-dimensional space containing 100 points. We set a value of $\sigma=1$ for the kernel matrix parameter in Eq.~\eqref{Eq:K_ij}. DPPs lead to Voronoi cells that have comparable areas, whereas TPPs lead to cells that have either small areas (around clusters) or very large areas (in between clusters). The PPP cells are also inhomegeneous but the level of size variation is smaller than that in the TPP sample.} \label{fig:vd}
\end{figure*}

\begin{figure}[b]
\centering
\includegraphics[scale=0.39]{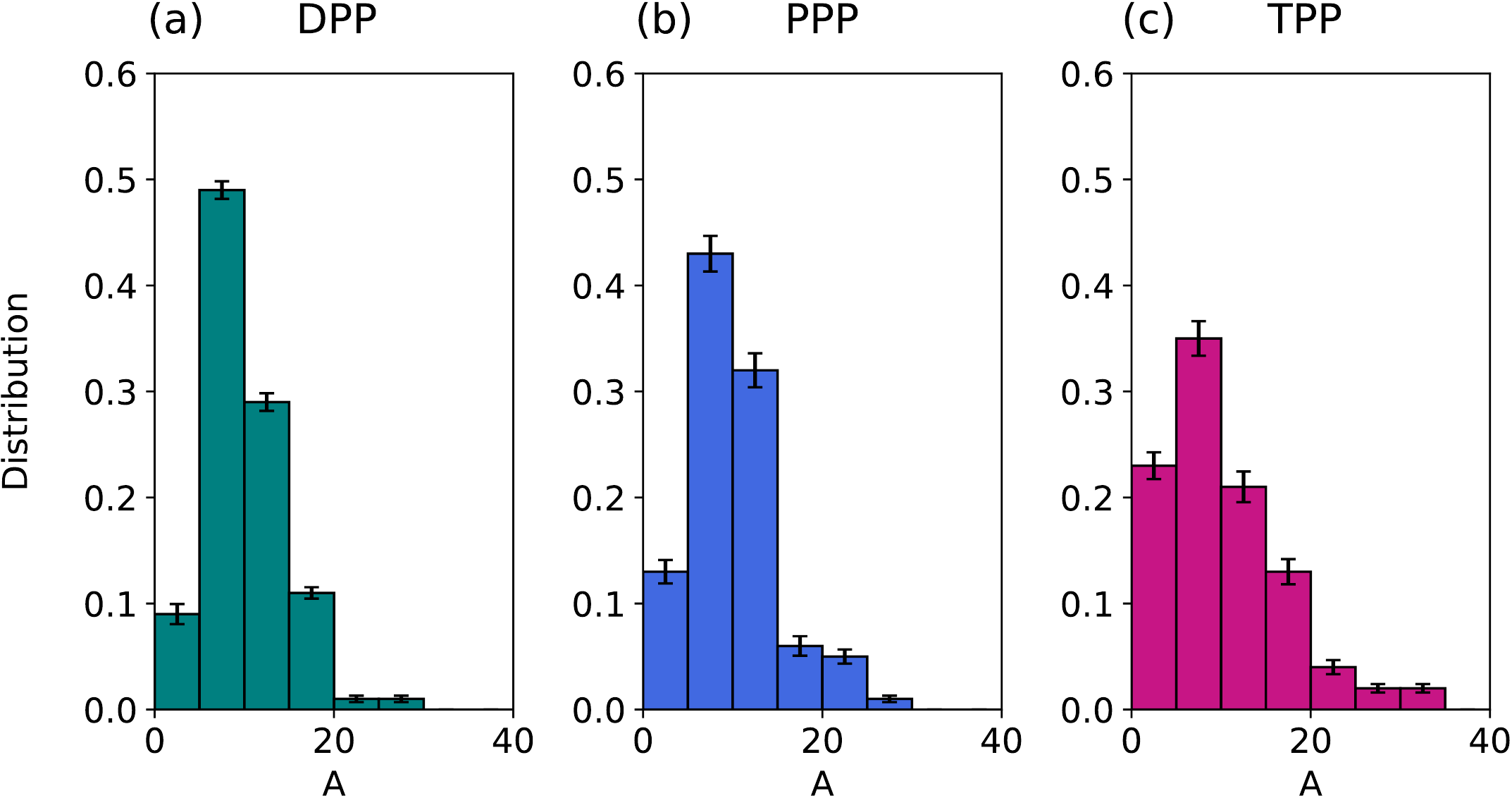}
\caption{Normalized distributions of the Voronoi cell areas for (a) determinantal, (b) Poisson, and (c) Torontonian point processes in a two-dimensional space containing 100 points. We set a value of $\sigma=1$ for the kernel matrix parameter in Eq.~\eqref{Eq:K_ij}. The statistics were taken over ten independent samples and the error bars correspond to one standard deviation. DPPs have a high peak in the distribution, indicating that the Voronoi cells have roughly the same area since the points are scattered evenly. For TPPs, there is a significantly large probability of small cell areas, which is to be expected when the patterns form several clusters of nearby points. The PPP distribution reflects the intermediate level of inhomogenity in the PPP cell sizes, compared to the DPP and TPP samples.}\label{fig:av}
\end{figure}

We now investigate the general properties of GBS point processes. Following the discussion in Section~\ref{Sec:PP}, the correlation function of an HPP is defined as 
\begin{equation} \label{rhoh}
\rho_n(r_1, \ldots, r_{2n}) = \text{Haf}[K_{r_i, r_j}],
\end{equation} 
where $2n$ refers to the number of points generated by the HPP. Similarly, the 2-point correlation function is
\begin{equation} \label{Eq:rhot}
\rho_2(r_1, r_2) = \text{Haf} 
\begin{bmatrix}
K_{r_1,r_1} & K_{r_1,r_2} \\
K_{r_2,r_1} & K_{r_2,r_2}
\end{bmatrix} =
K_{r_1,r_2}.
\end{equation}
According to Eq.~\eqref{Eq:rhot}, when the kernel matrix is constructed to quantify the similarity between the points, HPP selects pairs of similar points with higher probability. Note that $K_{r_1,r_2}\geq 0$ for all valid kernel matrices. This indicates that, as expected, HPPs sample points that are clustered together, i.e., are more similar, with higher probability. The same clustering property also holds for TPPs, since the coarse-graining that maps HPPs to TPPs leaves the interaction between neighbouring points unaffected. Implementing point processes with this form of collective clustering can be done natively with GBS and, in special cases, with quantum-inspired methods. In this section, we explore the properties of the GBS point processes by studying the role of different types of state spaces and kernel matrices in the context of example applications.

\begin{figure*}
\includegraphics[scale=0.12]{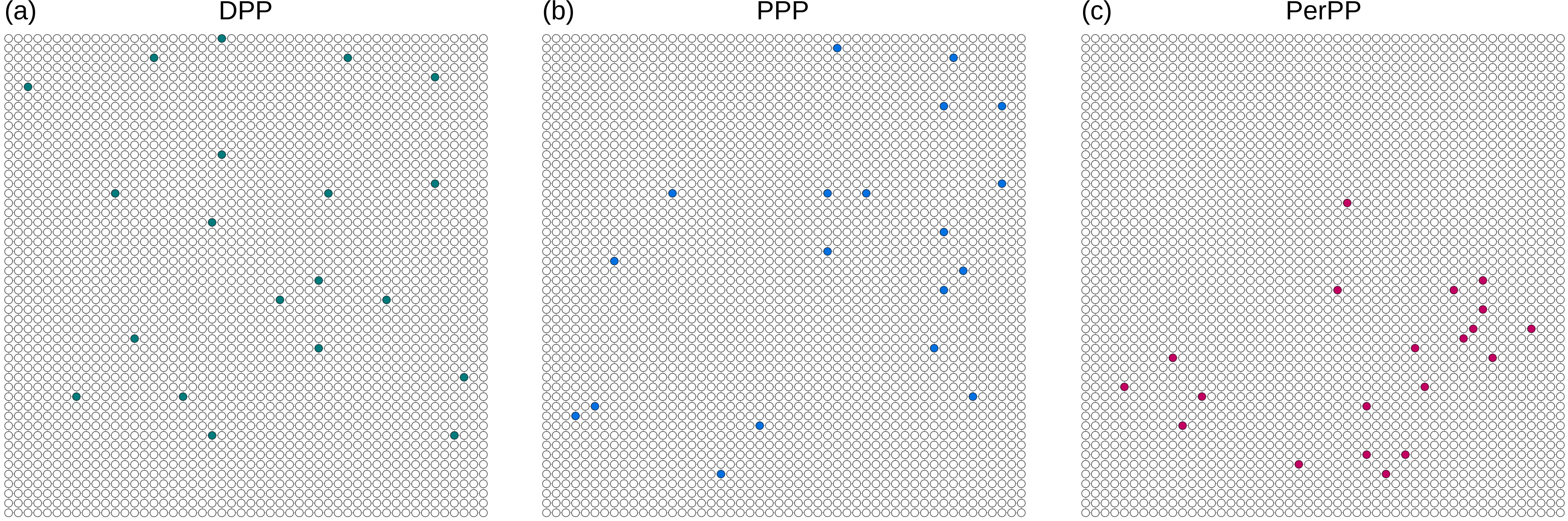}
\centering
\caption{Samples generated with (a) determinantal, (b) Poisson, and (c) GBS-inspired permanental point processes in a two-dimensional space containing 2500 points. DPP samples have points that are scattered and spread out in space. PPPs treat all patterns uniformly, so both clustering and repulsion are typically present. For the GBS-inspired PerPP, sample points are likely to occur in clusters of nearby points.}\label{fig:dpti}
\end{figure*}

\vspace{-0.4cm}

\subsection{Homogeneous state spaces}

We study spatial point processes where the state space is a set of points distributed uniformly in a two-dimensional space and the kernel matrix has elements given by \citep{Li2015Statistical}
\begin{equation} \label{Eq:K_ij}
K_{i,j} = e^{-\|\bm{r}_i-\bm{r}_j\|^2/\sigma^2},
\end{equation}
where $\bm{r}_i=(x_i,y_i)$ is the coordinate vector of the $i$-th point and $\sigma$ is a parameter of the model. In this scenario, similarity is determined in terms of spatial proximity: points that are close to each other are assigned large entries in the kernel matrix, with similarity decaying exponentially with distance. Now, we compare and analyze the statistical properties of the patterns that appear when employing determinantal, Poisson, and Torontonian point processes. Here and throughout the rest of the paper, DPPs are implemented using the algorithm of Ref.~\cite{kulesza2012determinantal}, and TPPs are implemented by employing the GBS simulation algorithm of Ref.~\cite{quesada2018gaussian}. As discussed previously, because the points generated with the PPP are distributed uniformly, both clustered and dispersed groups of points are equally likely to be observed. Conversely, the DPP point patterns are scattered over the whole state space, while the TPP points form well-defined clusters. Fig.~\ref{fig:pdh} illustrates typical samples from each of these point processes, which reflect their expected behaviour. More examples of such point patterns are provided in the Appendix. \\

Numerical evidence of the clustering properties of TPPs can be obtained by analyzing the distributions of the nearest-neighbour distances, $N(r)$, which characterize the probability of finding the closest neighbour of a point at a distance $r$. In Fig.~\ref{fig:nnd}, we report the empirical frequencies of nearest-neighbour distances for the determinantal, Poisson, and Torontonian point processes over 10 samples. The $N(r)$ for the TPP has a peak at the smallest possible distances between neighbours, showing that a large fraction of the points generated by the TPP have at least one neighbour in the closest possible position on the discrete state space. The $N(r)$ curve drops significantly for larger distances due to the small probability of observing scattered points in the TPP samples. The $N(r)$ obtained for the DPP has a peak at a relatively large distance because the points generated by a DPP repel each other. The PPP $N(r)$ is more uniform, compared to the TPP one, due to the comparable probability of finding points at small and large distances from each other.  \\

These features are also reflected in the Voronoi diagrams of the samples, which are shown in Fig.~\ref{fig:vd}. The majority of the cells in the DPP diagram have similar sizes due to the spread-out distribution of the points in the state space. In the PPP diagram, the homogeneity of the cell sizes decreases compared to the DPP, and in the TPP diagram, both very small and very large cell sizes are observed due to the appearance of point clusters. The normalized distributions of the areas of the Voronoi cells computed for 10 samples obtained from these point processes are shown in Fig.~\ref{fig:av}. These distributions provide further numerical evidence for the lower homogeneity of the TPP Voronoi cell sizes compared to DPP and PPP results, again due to the clustering of the points in TPPs.\\

Classical simulation of TPPs is generally intractable due to the computational hardness of calculating Torontonians of arbitrary kernel matrices. However, since the kernel matrix of spatial point processes is positive semidefinite, the methods explained in Section~\ref{Sec:GBSi} for sampling thermal states allow the application of a quantum-inspired spatial point process for large state spaces. A typical sample generated with the quantum-inspired algorithm for permanental point processes (PerPPs) is presented in Fig.~\ref{fig:dpti} for a larger state space containing 2500 points, along with the corresponding DPP and PPP samples. The PerPP indeed generates clustered point patterns analogous to the TPP ones. This makes quantum-inspired point processes the preferable method for efficient modeling of clustered point patterns for positive-semidefinite kernel matrices.

\subsection{Inhomogeneous state spaces}

\begin{figure}[b!]
\centering
\includegraphics[scale=0.3]{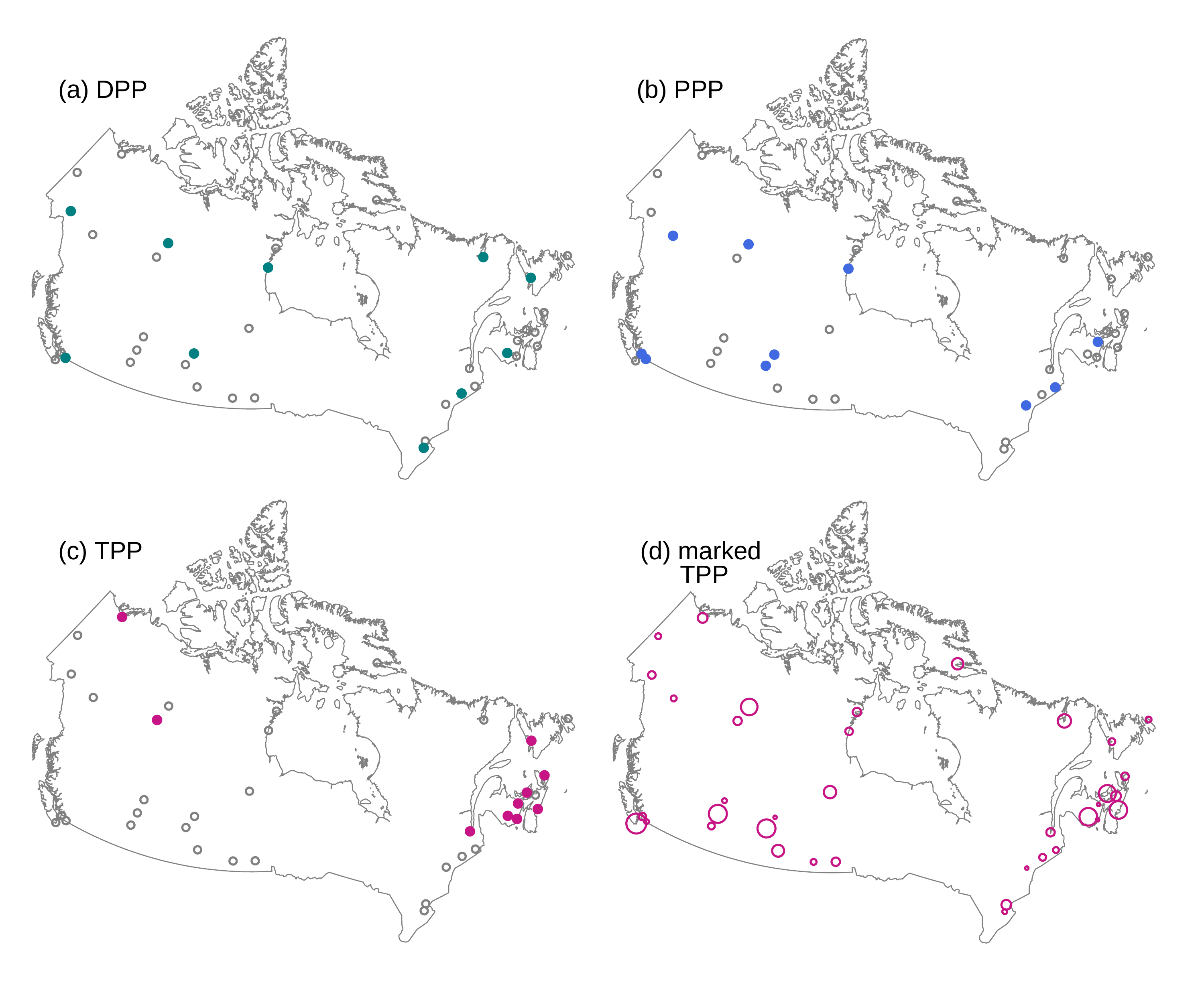}
\caption{Samples generated with (a) determinantal, (b) Poisson, and (c) Torontonian point processes in a two-dimensional space containing 37 Canadian cities. The parameter $\sigma$ in the kernel matrix in Eq.~\eqref{Eq:K_ij} was set to 1. The sizes of the circles in panel (d) indicate the number of times each city appeared in 100 TPP samples. In this setting, TPPs can help identify cities that have other large cities nearby.}\label{fig:ca}
\end{figure}

In the case of an inhomogeneous state space, i.e., a state space where points are not evenly separated, TPPs with a kernel matrix defined in Eq.~\eqref{Eq:K_ij} sample from regions containing many nearby points with high probability. This is in contrast with previous examples where clusters were equally likely to appear in any regions of the state space. In Fig.~\ref{fig:ca}, samples generated from the three point processes for an inhomogeneous state space formed based on the locations of 37 Canadian cities are presented. The state space is formed with the locations of the three most highly-populated cities in each province or territory. The TPP sample assigns a high probability to clusters of cities that are nearby, therefore identifying provinces whose largest cities are located near each other. The PPP sample also contains small clusters due to the random nature of the process. However, the points generated by DPP are fairly scattered. The preference of TPP for sampling from dense regions can be used to mark the points of a state space by assigning them with values that specify the number of times they appear among several TPP outputs. The magnitude of these marks is a measure of the closeness of a point to its neighbours. An example of such marked points is shown in Fig.~\ref{fig:ca}(d), where the number of times that each city appears in a TPP output was aggregated for 100 runs. The resulting marks are typically larger for cities located in dense regions as they tend to appear more frequently in the TPP samples.

\begin{figure*}
\includegraphics[scale=0.45]{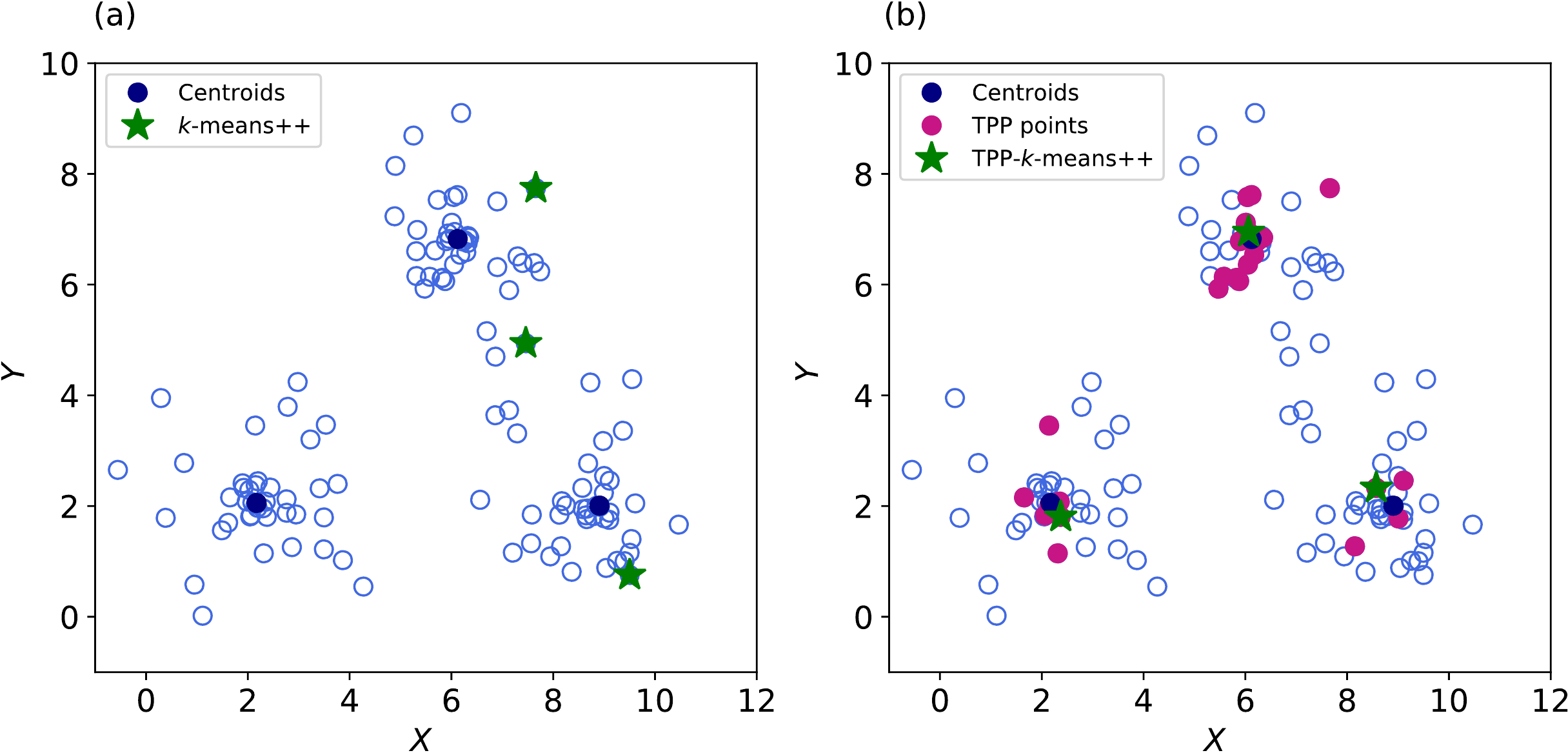}
\caption{(a) A dataset containing 120 points partitioned into 3 clusters with the initial seeds generated by the $k$-means++ method. The actual cluster centroids are also shown. (b) Application of TPP to the same dataset provides points that are close to the actual cluster centers. The initial seeds generated by the $k$-means++ method applied to the TPP points are also shown.}\label{fig:cls}
\end{figure*} 

The appearance of clustered points in the TPP outputs of inhomogeneous state spaces is a feature that can be used in different applications. For instance, TPPs could be implemented to increase the efficiency of classical clustering algorithms. Several problems in data mining and machine learning require the classification of unlabeled data into separate groups or categories. A clustering algorithm that is commonly used in this context is the $k$-means++ method, which partitions a number of data points into $k$ clusters such that each data point belongs to the cluster with the closest mean \cite{Arthur2007kmeans}.\\

In the $k$-means++ algorithm, the initial assignment of the cluster means is arbitrary and the centers selected initially can be far from the actual cluster means. We have already observed that the application of TPPs to such state spaces generates point patterns that are clustered at the high density regions. These points are likely to be close to the actual cluster means and therefore can be used as candidates for selecting better initial means in the $k$-means++ algorithm. We apply this procedure on a dataset containing 120 points partitioned into three clusters. The clusters were generated by choosing three cluster centers randomly and then generating 40 points from Gaussian distributions centered at each of these initial centers. The dataset and the cluster centers are shown in Fig.~\ref{fig:cls}(a).\\

Application of TPP to this dataset provides points that are close to the actual cluster means. Examples of such TPP point patterns are shown in Fig.~\ref{fig:cls}(b). The application of TPP transforms the original dataset into a smaller set of points that are well-separated and, more importantly, fairly concentrated around the cluster centers. As a result, the $k$-means++ algorithm applied to the points generated by TPP provides, with a higher probability, a set of initial centroids that are very close to the actual cluster means. An example of such initial centroids, selected by the the $k$-means++ algorithm from the TPP points, is shown in Fig.~\ref{fig:cls}(b). The combination of the TPP and $k$-means++ methods increases the probability of initiating the clustering procedure from a better set of initial seeds, which can potentially make the whole clustering processes more efficient.

\subsection{Cluster location}

Kernel matrices have a central effect on the properties of matrix point processes. In the examples considered so far, the elements of the kernel matrices were functions of the Euclidean distance between the points of a spatial state space. However, the components of the kernel matrix can be designed to represent additional features. For instance, in a homogeneous state space where the points are evenly distributed, TPP outputs contain local clusters in regions of the state space without any preference for where these clusters are located. To introduce control over the location of clusters, the kernel matrix can be rescaled to favor the appearance of points in hotspot regions.\\

\begin{figure} [b]
\includegraphics[scale=0.6]{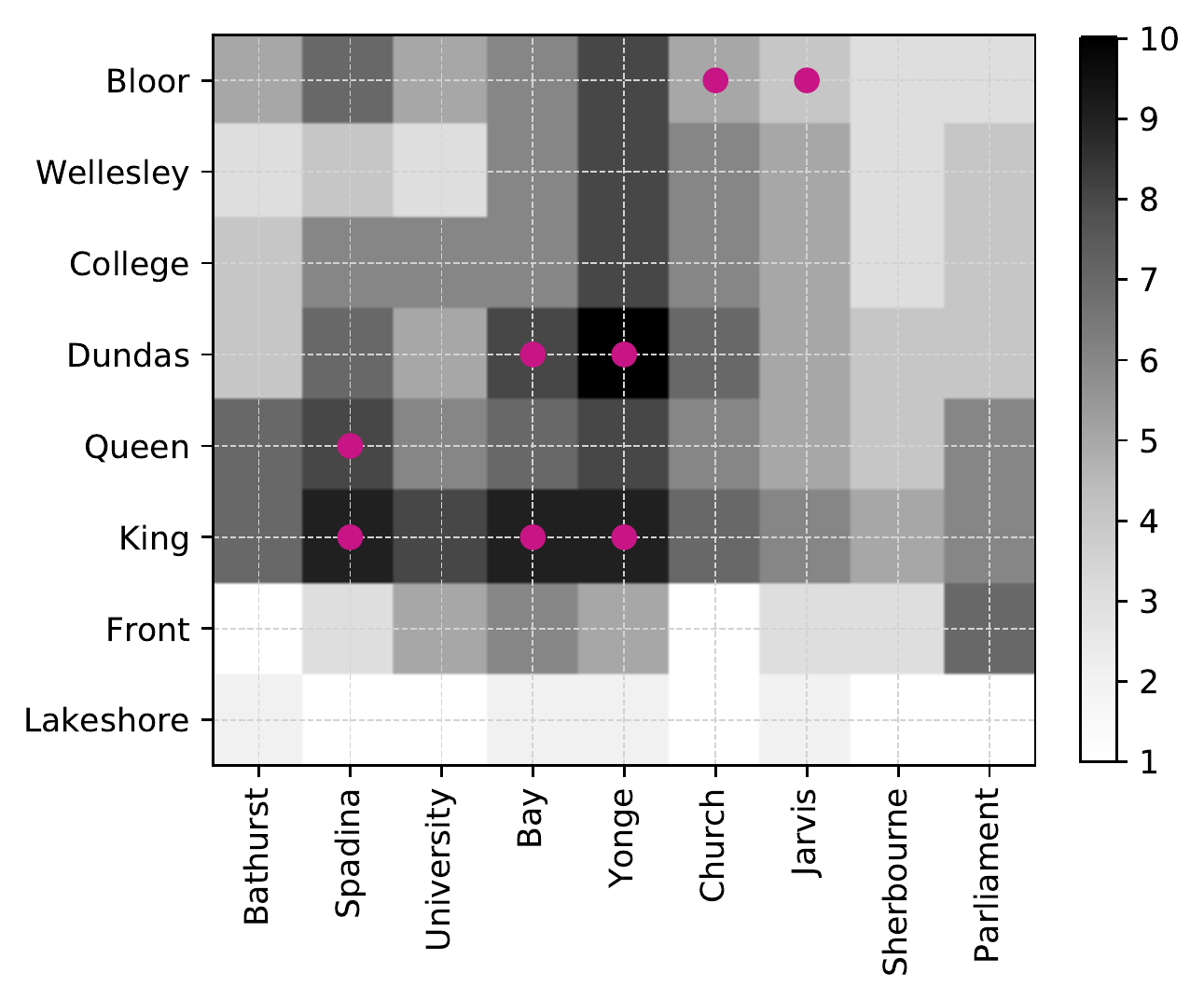}
\centering
\caption{A TPP point pattern sampled with a kernel matrix that has been rescaled according to the estimated population density of each major intersection in downtown Toronto, which was assigned a density on a scale of 1 to 10. The density is illustrated by the shade of the squares centered at the intersections, with darker colors corresponding to higher density.}\label{fig:rd}
\end{figure}

\begin{figure*}[t!]
\includegraphics[scale=0.42]{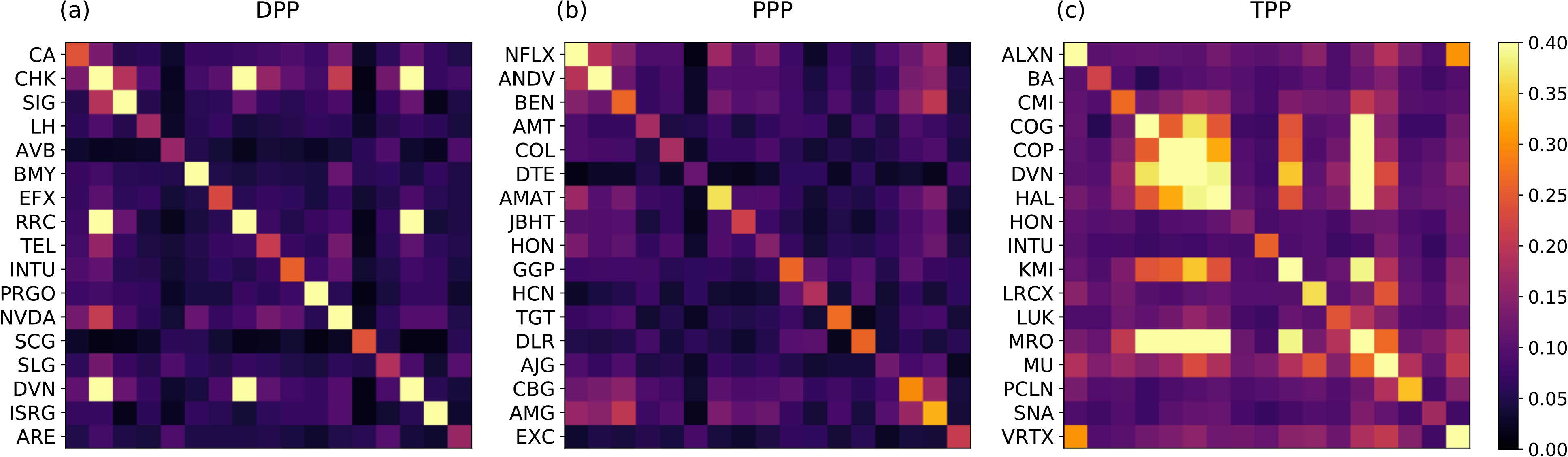}
\centering
\caption{Typical samples of correlation matrices of stocks selected from the S\&P 500 index collection by applying (a) determinantal, (b) Poisson, and (c) Torontonian point processes. The kernel matrix used to implement the point processes is the correlation matrix constructed from the market data of the stocks of the S\&P 500 index. The point process samples are subsets of stocks, here shown by their stock symbols. For each subset, we display the corresponding correlation matrix. Lighter points correspond to large entries of the correlation matrix. The TPP sample contains stocks that are highly correlated while the DPP and PPP samples contain stocks with lower levels of correlation.}\label{fig:cm}
\end{figure*}

One concrete method of adding control over the location of point clusters is to assign a density to each point in state space, resulting in a density vector $\bm{\lambda}=(\lambda_1, \lambda_2,\ldots, \lambda_m)$, with $\lambda_i\geq 0$. In this case, the kernel matrix of Eq.~\eqref{Eq:K_ij} can be adapted to
\beq \label{Eq:K_lambda}
K_{i,j} = \lambda_i\lambda_je^{-\|\bm{r}_i-\bm{r}_j\|^2/\sigma^2}.
\eeq
We apply a TPP with this kernel matrix to generate samples in a homogeneous state space defined with discrete points distributed evenly in a uniform grid, where each point represents the intersection of two major streets in the downtown area of the city of Toronto. The background population density at each intersection was approximately estimated from our own experience, and included in the kernel matrix of equation~\eqref{Eq:K_lambda}. The variation in the distances between the neighboring intersections was neglected for simplicity. An example of this statistical model is shown in Fig.~\ref{fig:rd}, where we illustrate a typical sample of points. The TPP samples generated with the resulting kernel matrix produce points that are clustered in more congested areas of the city. More sophisticated models could be built from such GBS point processes by fine-tuning the kernel matrix and introducing more elaborate rules to generate 
samples that can be used to model different types of events occurring in a region. 

\subsection{Correlation between points}

The state space of a matrix point processes can be designed to represent objects that are not necessarily actual points in a physical space. In fact, the state space can correspond to any collection of items. Additionally, kernel matrices may represent more general forms of correlations between points, not just those due to spatial proximity. This opens up new fields of application for generating samples in abstract spaces. For instance, a central problem in financial optimization is to select groups of stocks and bonds either for investment or to be combined into new financial products. To reduce investment risk, it is important to avoid buying correlated assets whose prices fluctuate in synchrony, in which case it is helpful to identify them \cite{Leon2017Clustering}. Additionally, finding groups of correlated assets can reveal market trends that may be otherwise difficult to anticipate.\\

GBS point processes can be used to identify correlated stocks by setting the correlation matrix of the stocks as the kernel matrix of the point process. To construct the correlation matrix, we take a vector of returns $\bm{R}_j=(R_{1,j}, R_{2,j}, \ldots, R_{n,j})^T$, where $R_{i,j}$ is the return of stock $i$ on the $j$-th day. The correlation matrix $\Sigma$ is given by
\begin{align}
\Sigma = \frac{1}{n}\sum_{j=1}^n \bm{R}_j\bm{R}_j^T.
\end{align}
The point patterns generated by TPPs applied with such kernel matrices are expected to contain stocks with higher levels of correlations, i.e., the clusters correspond to collections of correlated stocks.\\ 

Here we apply TPP to stocks of the S\&P 500 index, constructing a correlation matrix from publicly-available pricing data for the stocks 
comprising the S\&P 500 stock index during the five-year period 2013-2018~\cite{nugent2018kaggle}. Not all stocks remain in the index during this time period, so we focus on the 474 stocks that do. We use this correlation matrix directly as the point process kernel matrix. In this setting, point patterns correspond to subsets of stocks, which we illustrate in terms of their respective covariance matrix. \\

Typical samples from DPP, PPP, and TPP are shown in Fig.~\ref{fig:cm}. A sample in this context is a subset of stocks, illustrated by the corresponding correlation matrix. Comparison of the TPP result with those containing the same number of stocks selected by DPP and PPP indicates the ability of TPP in detecting stocks with higher levels of correlation. For example, the four stocks that form a cluster on the top left corner of Fig.~\ref{fig:cm} (c) correspond to Cabot Oil \& Gas (COG), ConocoPhillips (COP), Devon Energy (DVN), and Halliburton (HAL), all of which are companies involved in petroleum exploration and therefore expected to be highly correlated. It is noted that clustering algorithms have been used in portfolio optimization and can provide more stable investment strategies compared to conventional techniques \cite{Leon2017Clustering}. Similarly, TPP could be used in this context to detect correlations that might be even hidden from classical clustering techniques.

\section{Conclusions}
We have proposed an application of quantum computing to statistical modeling by introducing a class of point processes that can be implemented with special-purpose photonic quantum computers. These point processes are generally intractable to implement with conventional methods but, as we have shown, they can be efficiently implemented with GBS devices. For models with positive semidefinite kernel matrices, we have developed fast quantum-inspired algorithms whose runtime is quadratic in the size of the state space. This includes an efficient algorithm for permanental point process which did not exist previously in the literature. Our results open up the possibility of a wider application of point processes that generate clustered data points, which were previously less explored due to the challenges in their implementation.\\

Point processes can be employed to provide insights into stochastic phenomena of interest, represent patterns with desired properties, or help with the identification of specific structures. Besides these general applications, the GBS point processes developed here can be implemented in many other different scenarios. Here we explored some of the potential applications of the GBS point processes to illustrate their versatility and usefulness. Further work is required to fully understand and quantify the advantages of using GBS in these contexts, but our results already give an insight of the scope of this photonic quantum technology. For example, kernel matrices play a key role in determining the properties of the generated point patterns. We focused our attention on kernel matrices that represent the similarity between the points, a specific choice that was adopted because clustering of the resulting point patterns was important for the applications we considered. However, many other options are possible: kernel matrices can reflect differences between points, they can represent graphs in the form of adjacency matrices as in Refs.~\cite{bradler2018gaussian, arrazola2018using, arrazola2018quantum, bradler2018graph, schuld2019quantum}, or they can be trained from data to produce desired patterns. \\ 

The fundamental connection between point processes and GBS can also be harnessed from another perspective. The statistical features of the point processes that have been analyzed analytically and numerically here can be used as indicators for validating the correctness of physical GBS machines. For instance, a GBS point process implemented with the kernel matrices used here will result in point patterns with enhanced aggregation of points. Accordingly, any GBS device programmed according to such matrices should output clustered point patterns, a feature that if verified in an actual implementation can be used as an initial signal of the proper functioning of the device. \\ 

From a fundamental perspective, our work brings the connection between statistical modeling and physical systems full circle: not only can mathematical models be used to simulate natural processes, physical systems themselves can be engineered and programmed to implement abstract models. Remarkably, this is ultimately possible by controlling the behavior of fundamental particles --- photons --- and mediating their interactions via macroscopic matter.\\

\section*{Acknowledgements}
We thank David Duvenaud, Luke Helt, Alain Delgado, Zeyue Niu, Rafal Janik, Paul Finlay, Smriti Shyamal and Christian Weedbrook for valuable discussions. We also thank Andrey Goussev for providing the picture in Fig.~\ref{fig:bee}.

\setcounter{figure}{0}    
\renewcommand\thefigure{A.\arabic{figure}}
\begin{figure*}
\includegraphics[scale=0.85]{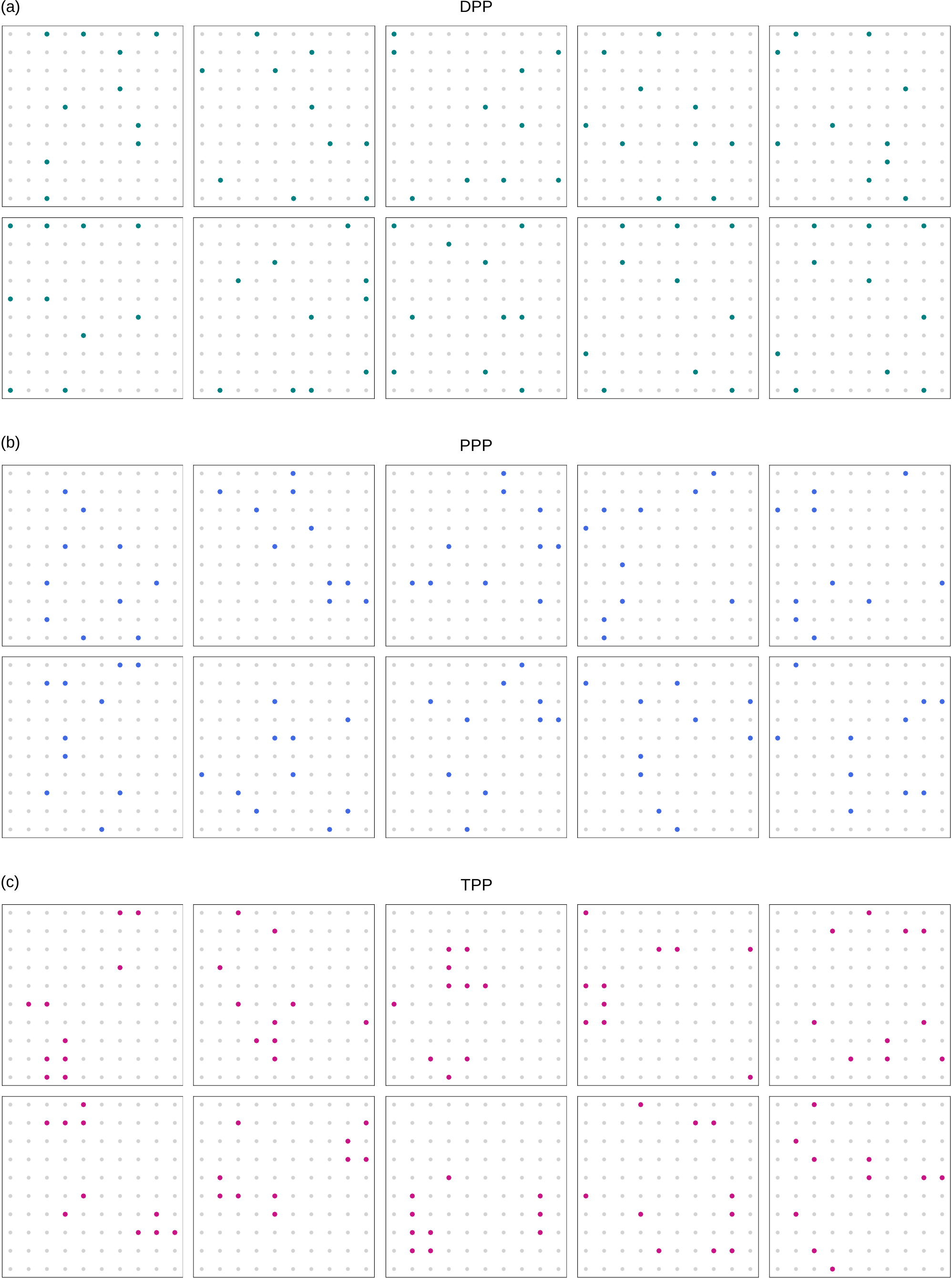}
\centering
\caption{Samples generated with determinantal, Poisson, and Torontonian spatial point processes in a two-dimensional space containing 100 points. DPP was set to generate 10 points for each sample. The patterns generated with PPP and TPP usually contain different number of points but only those samples that contained 10 points are shown here for consistency.}\label{fig:apx}
\end{figure*}

\section*{Appendix}
Point patterns generated with determinantal, Poisson, and Torontonian point processes in a state apace containing 100 points, spread evenly in a two dimensional space, are presented in Fig.~\ref{fig:apx}. These samples are provided to visually illustrate the typical features of the point patterns and complement the results presented in Section~\ref{Sec:prop}. Inspection of the patterns demonstrate that the points generated by DPP are scattered and spread out in space, the PPP patterns contain both clustering and repulsion, and clustering of the points is more probable in TPP samples. 
\vfill\null
\vspace{13cm}

\bibliographystyle{apsrev}
\bibliography{HPP_references}

\end{document}